\documentclass{article}  
\usepackage{breckenr2005}
\frompage{000} \topage{000}                                              
\def\rightmark{Triaxiality, chirality, and $\gamma$-softness.}
\def\leftmark{K.~Starosta et al.}
 
\title{Triaxiality, chirality, and $\gamma$-softness} 
\authors{
{K.Starosta$^{1,2}$, M.A.Caprio$^{3}$, T.Koike$^{4}$, 
R.Kr\"ucken$^{5}$, and C. Vaman$^{2}$}\\[2.812mm]
{\normalsize
\hspace*{-8pt}$^1$ Department of Physics and Astronomy,\\
Michigan State University, East Lansing, Michigan 48824, USA\\[0.2ex] 
\hspace*{-8pt}$^2$ National Superconducting Cyclotron Laboratory,\\ 
East Lansing, Michigan 48824, USA\\[0.2ex] 
\hspace*{-8pt}$^3$ Wright Nuclear Structure Laboratory,\\ 
Yale University, New Haven, Connecticut 06520, USA\\[0.2ex]
\hspace*{-8pt}$^4$ Department of Physics,\\
Tohoku University, 980-8578 Sendai, Japan\\[0.2ex]
\hspace*{-8pt}$^5$ Physik Department E12,\\
Technische Universit{\"a}t M{\"u}nchen, D-85748 Garching, Germany\\[0.2ex]
}}
 
\abstract{Current work explores the impact of $\gamma$-softness on
partner bands built on the $\pi$h$_{11/2}$$\nu$h$_{11/2}$
particle-hole configurations in triaxial odd-odd nuclei.  The results
of calculations conducted using a core-particle-hole coupling are
presented. The model Hamiltonian includes the collective core, the
single-particle valence nucleons, and separable quadrupole-quadrupole
interactions.  The Kerman-Klein method was applied to find eigenstates,
which provided a convenient way for exploring core effects.
Calculations were made for triaxial cores with various
$\gamma$-softness using the General Collective Model keeping the
expectation value for the triaxiality parameter fixed at
$\langle\gamma \rangle =30^\circ$. The degeneracy in the
$\pi$h$_{11/2}$$\nu$h$_{11/2}$ bands results from the calculations for
the $\gamma$-rigid core but is lifted for the $\gamma$-unstable core.}

 
\begin{document}
 
\maketitle
\setcounter{page}{1}

\begin{figure}[t]
\vspace*{-0.9cm}
                 \insertplot{ 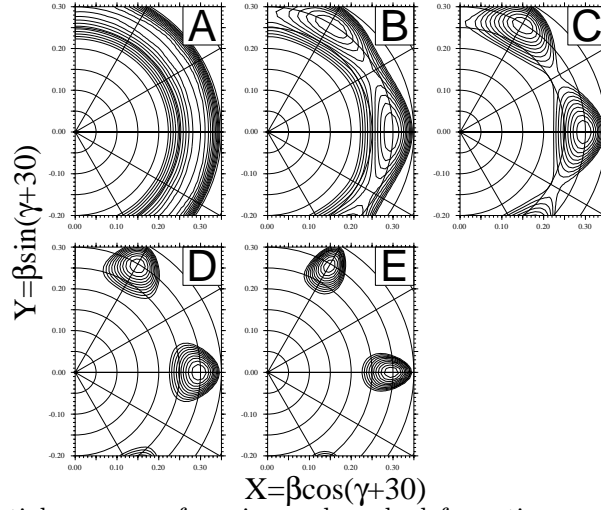}
\vspace*{-0.8 cm}
\caption[]{Potential energy surfaces in quadrupole deformation
coordinates $\beta$ and $\gamma$ for triaxial cores of varying
$\gamma$-softness ($\langle \gamma^2 \rangle$-$\langle \gamma
\rangle^2$) with the expectation values of other deformation
parameters such as $\langle \beta \rangle$ , $\langle \beta^2 \rangle$
and $\langle \gamma \rangle$ constrained to be nearly identical.}
\label{fig1}
\end{figure}

\begin{figure}
\vspace*{-.0cm}
                 \insertplot{ 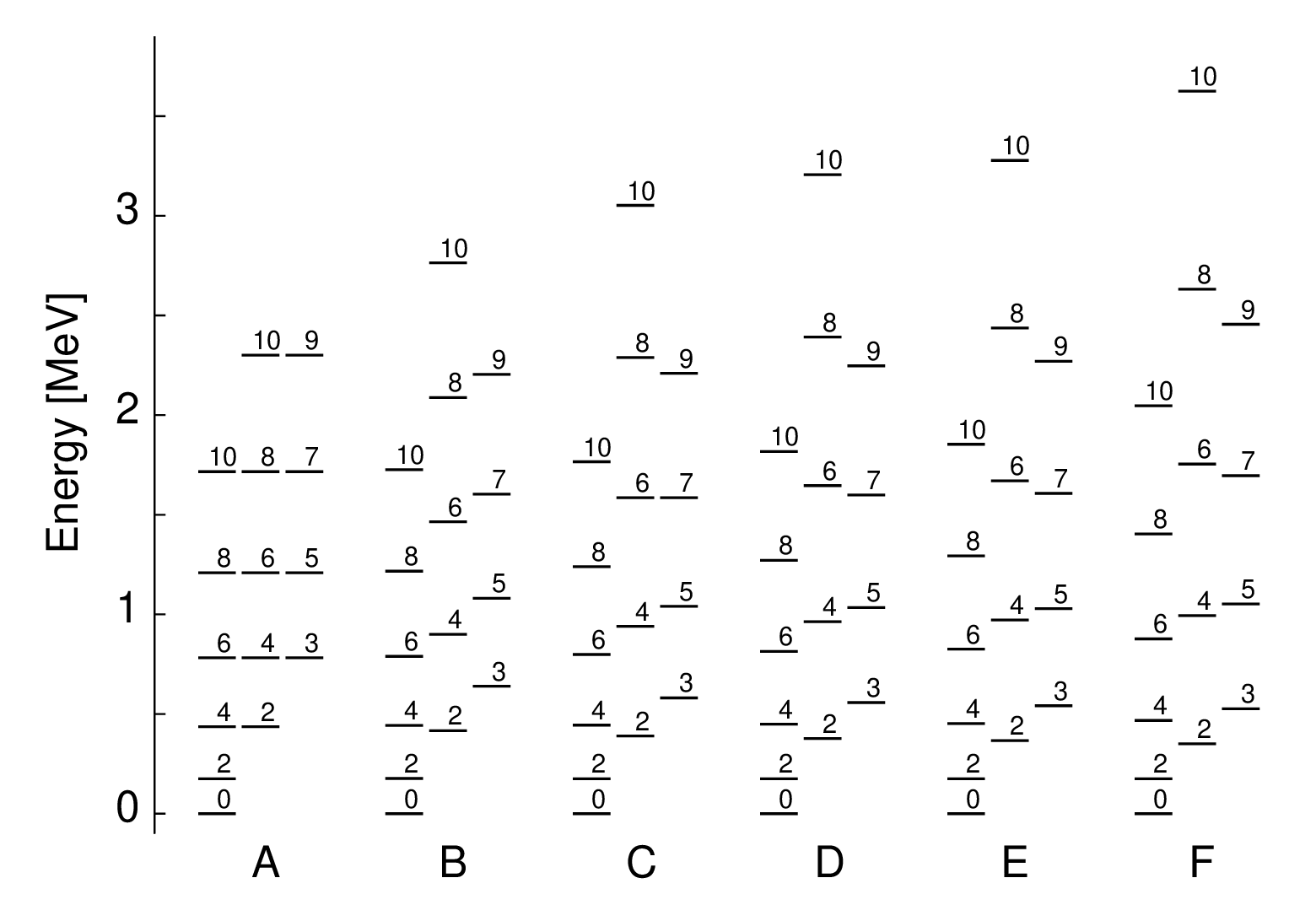}
\vspace*{-0.9 cm}
\caption[]{Energy levels which result from the diagonalization of the Bohr
Hamiltonian following the General Collective Model of Ref.\ \cite{gra87} 
for ({\bf A}-{\bf E}) potential energy surfaces shown in Fig.\ \ref{fig1}  
and ({\bf F}) for the rigid triaxial rotor of Ref.\ \cite{dav58}.    }
\label{fig2}

\vspace*{-0.0cm}
                 \insertplot{ 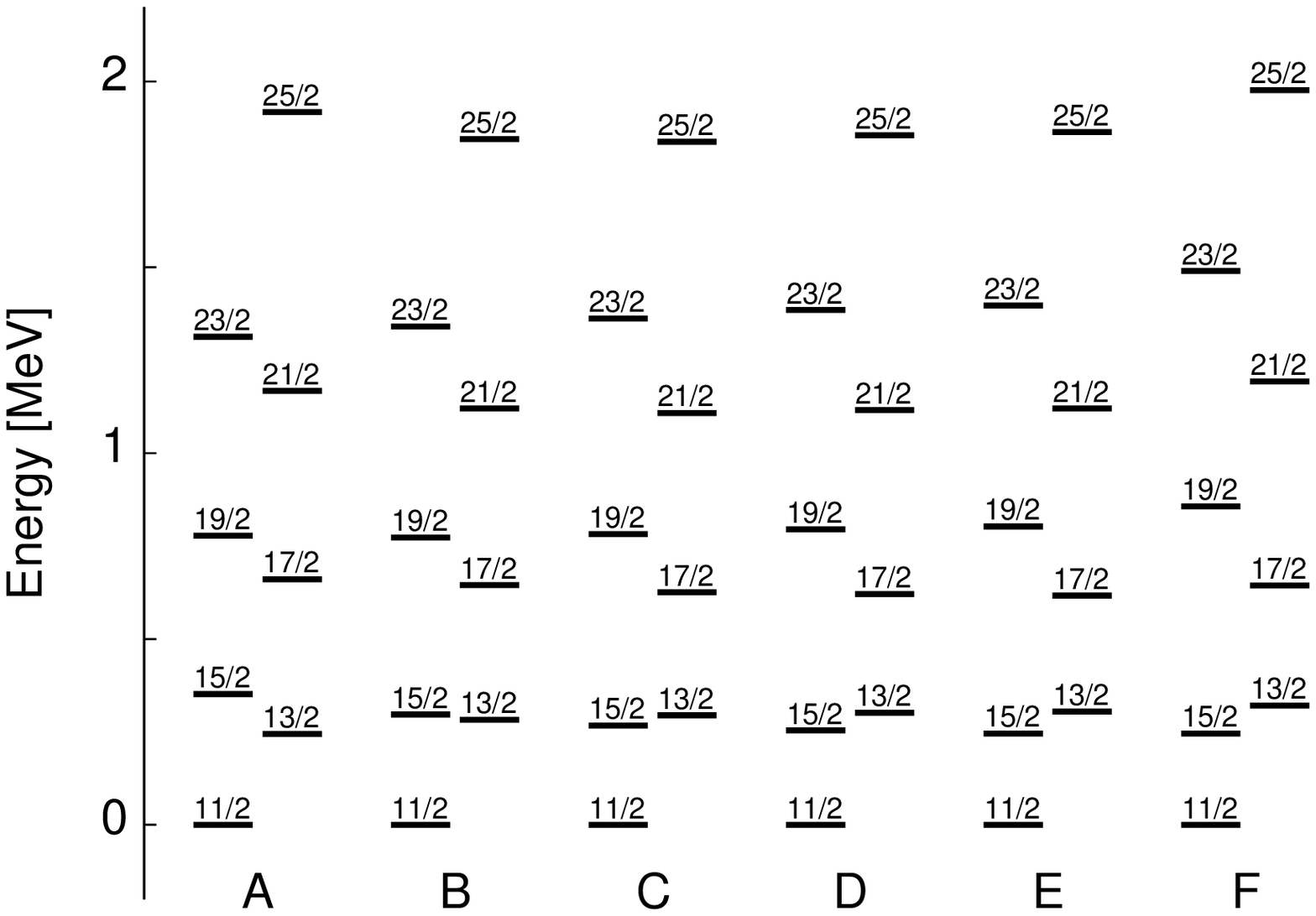}
\vspace*{-0.5 cm}
\caption[]{Energy levels in yrast bands of odd nuclei as resulting
from a valence h$_{11/2}$ odd-particle or odd-hole coupling to the
cores {\bf A}-{\bf F}.  The calculations follow the Kerman-Klein
method discussed in Ref.\ \cite{sta02,koi03}.}
\label{fig3}
\end{figure}

While the impact of quadrupole deformation on the structure of nuclei
away from the closed shell is well recognized\ \cite{rag95}, the
existence of static triaxial shapes continues to be a subject of a
long standing debate. The rigid triaxial- \ \cite{dav58} and
$\gamma$-unstable \cite{wil56} rotor models yield very similar
predictions for the yrast states in nuclei with even number of protons
and neutrons (even nuclei) as illustrated by the calculations
presented in Figs.\ \ref{fig1} and\ \ref{fig2}.

Figure\ \ref{fig1} shows the potential energy surfaces in quadrupole
deformation coordinates $\beta$ and $\gamma$ for triaxial cores of
varying $\gamma$-softness ($\langle \gamma^2 \rangle$- $\langle \gamma
\rangle^2$) with the core labeled as {\bf A} being $\gamma$-unstable,
and cores labeled {\bf B}-{\bf E} becoming more $\gamma$-rigid; the
potential energy surface for the rigid triaxial core labeled below as
{\bf F} is not shown in Fig.\ \ref{fig1} since it corresponds to the
Dirac's $\delta$ function at $\gamma=30^\circ$, $90^\circ$, and
$120^\circ$. The above potential energy surfaces were constructed to
differ by the $\gamma$-softness parameter only, meaning that the care
was taken to assure that the expectation values of other deformation
parameters such as $\langle \beta \rangle$ , $\langle \beta^2 \rangle$
and $\langle \gamma \rangle$ are nearly identical.

These potential energy surfaces were used to solve the collective Bohr
Hamiltonian following the General Collective Model of Ref.\ \cite{gra87}
with the computer code described in Ref.\ \cite{tro91}.  The resulting
energy levels are compared in\ \ref{fig2}. It can be observed in this
figure that the yrast states, shown for each core in the corresponding
column on the left hand side, are not effected by the
$\gamma$-softness.  Possible signatures of $\gamma$-softness or
$\gamma$-rigidity from these calculations involve non-yrast states and
are likely to be perturbed by single particle degrees of freedom which
are not taken into account in the model Hamiltonian. As a consequence,
conclusive results on stability of triaxial deformation in even nuclei
are hard to obtain and seems beyond the sensitivity of the current
experimental investigations.

The situation is not very different if the calculated yrast bands in
odd-mass nuclei (odd nuclei) are examined. Odd nuclei can be
considered as consisting of an even core with a coupled valence
quasiparticle. The results presented in Fig.\ \ref{fig3} for a
core-quasiparticle model calculations following the Kerman-Klein
method\ \cite{sta02,koi03} demonstrate essentially insignificant
dependence of the yrast state energies in odd nuclei on the
$\gamma$-softness of the {\bf A}-{\bf F} cores used for the coupling
with an unique-parity h$_{11/2}$ state. The results shown in Fig.\
\ref{fig3} are valid for a pure h$_{11/2}$ particle or hole coupling
due to the particle-hole symmetry discussed in Ref.\ \cite{mtv75}.

\begin{figure}
\vspace*{-1.0cm}
                 \insertplot{ 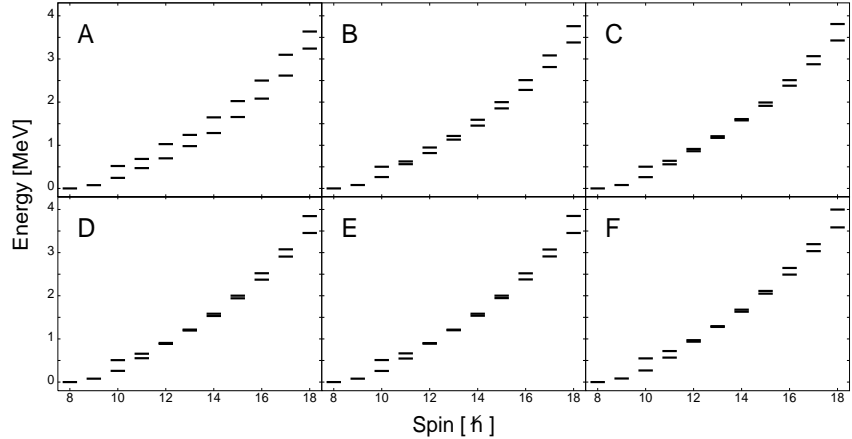}
\vspace*{-1.5 cm}
\caption[]{Energy levels in yrast and yrare bands of doubly-odd
nuclei as resulting from a valence odd-particle and odd-hole coupling
in the $\pi$h$_{11/2}$$\nu$h$_{11/2}$ configuration to the cores {\bf
A}-{\bf F}.  The calculations follow the Kerman-Klein method discussed
in Ref.\ \cite{sta02,koi03}.}
\label{fig4}
\end{figure}

As a surprise, therefore, may come the observation that particle-hole
coupling calculations which follow the model developed in Ref.\
\cite{sta02,koi03} for nuclei with odd number of protons and neutrons
(doubly-odd nuclei) show sensitivity of the yrast and near yrast
energy levels to the degree of $\gamma$-softness. This is illustrated
in Fig.\ \ref{fig4} presenting the two lowest energy states at a given
spin calculated for the $\pi$h$_{11/2}$$\nu$h$_{11/2}$ configuration
and the cores defined above. The separation between the states of the
same spin in the mid-spin range near $\sim$14$\hbar$ is observed to
decrease with decreasing $\gamma$-softness by nearly an order of
magnitude, from $\sim 400$ keV for the $\gamma$-unstable core {\bf A}
to $\sim 40$ keV for the $\gamma$-rigid core {\bf F}. The proposed
interpretation of this effect invloves a doubling of states resulting
from chiral coupling of angular momenta vectors for stable triaxial
core\ \cite{sta02} and lack of such doubling for the $\gamma$-soft
core for which the stable chiral geometry cannot be formed.

It should be pointed out here that the calculations presented in Fig.\
\ref{fig4} are generic and should be treated as qualitative rather than
quantitative.  This results from the fact that the parameters of the
model were not optimized to fit any particular nucleus. In addition
the core basis had to be truncated at spin 10$^+$ as shown in Fig.\
\ref{fig2}; this truncation is imposed by the General Collective Model
cores available to the authors. However, while the numerical values of
the energy separation between levels of the same spin in the mid-spin
range for doubly-odd nuclei may change depending on the details of the
calculations the correlation of the $\gamma$-softness and the energy
separation is expected to stay.  The study of particle-hole
configurations in doubly-odd transitional nuclei, therefore, can shed
light on the $\gamma$-stability of their triaxial cores.

As mentioned above, the small energy spacing between states of the same
spin in doublet bands built on the particle-hole configuration has
been proposed as a signature of formation of chiral geometry in
triaxial doubly-odd nuclei \ \cite{sta04}. Nuclear chirality is a
manifestation of spontaneous symmetry breaking\ \cite{fra01} resulting
from an orthogonal coupling of angular momentum vectors in triaxial
nuclei which minimizes the total energy of the system. For doubly-odd
nuclei three perpendicular angular momenta provided by the valence
particle, valence hole and the collective core rotation can form two
geometries of the opposite handedness; these two geometries are
related by the time reversal operator, which reverses an orientation
of each of the angular momentum components.

\begin{figure}
\vspace*{-1.0cm}
                 \insertplot{ 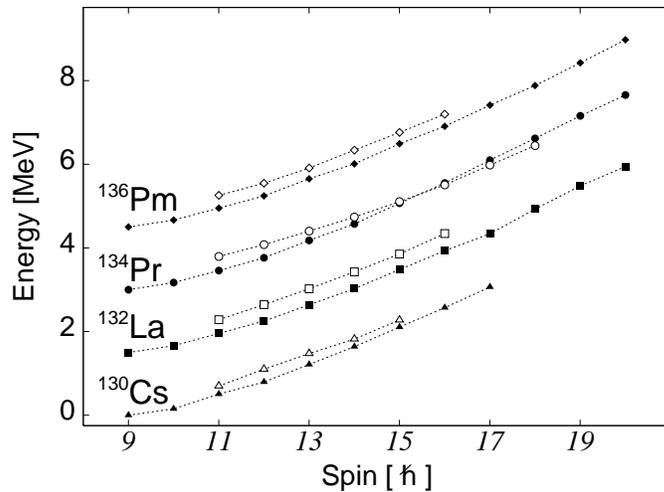}
\vspace*{-1.0 cm} 
\caption[]{Experimental energies for the
$\pi$h$_{11/2}$$\nu$h$_{11/2}$ partner bands in N=75 doubly-odd
isotones in the A$\sim$130 region. Bandhead energies are separated by
1.5 MeV for display. }
\label{fig5}
\end{figure}

Up to date a number of cases has been identified experimentally in
the A$\sim 130$ and the A$\sim 104$ regions with the doublet bands
built on the $\pi$h$_{11/2}$$\nu$h$^{-1}_{11/2}$ and the
$\pi$g$^{-1}_{9/2}$$\nu$h$_{11/2}$ configurations,
respectively. Figure\ \ref{fig5} shows the Energy vs. Spin plot for
four N=75 isotones discussed in Ref.\ \cite{sta01}. The experimental
data for the band's energetics seem to be in good qualitative
agreement with the calculations shown in Fig.\ \ref{fig4}. For the 
quantitative agreement, however, the calculations need to be performed
with more complete core basis and optimized set of model parameters.

One of the recent important observations related to the properties of
the doublet bands in the mass 130 region is a notable difference in
the electromagnetic transition rates between the doublet bands\
\cite{sta04}. It seems very interesting to examine the
$\gamma$-softness effects on the band's electromagnetic properties,
however, the core basis truncation in the current studies limits the
meaningfull comparison due to the fact that the electromagnetic
transition rates depend on the details of the wave function in more
sensitive way than the state energies.  Again, improved calculations
with extended core space are needed.

In conclusion, the presented core-particle-hole coupling calculations
indicate a correlation between the energy separation in doublet bands
built on the unique-parity intruder states and $\gamma$-rigidity of
the collective core in doubly-odd nuclei. Small energy separation
between states of the same spin in the medium-spin range results from
the calculations with a relatively $\gamma$-rigid core but this
degeneracy is lifted if the core is $\gamma$-unstable. Experimentally
there are two doubly-odd nuclei identified so far with small
separation between doublet bands: $^{104}$Rh in the mass $\sim 104$
region with the two 17$^-$ states built on the
$\pi$g$^{-1}_{9/2}$$\nu$h$_{11/2}$ configuration being 2~keV apart\
\cite{vam04} and $^{134}$Pr with the states at 15$^+$ and 16$^+$ built
on the $\pi$h$_{11/2}$$\nu$h$^{-1}_{11/2}$ configuration being less
than $\sim$50~keV apart\ \cite{sta01} . Further, improved calculations
are needed to examine a possible stable triaxiality of these
nuclei. These improved calculations should also be examined for
consistency of electromagnetic rates with the experimental data.

This work has been supported in part by the NSF 
grant number PHY01-10253 and the DOE grant number DE-FG02-91ER-40609.

\vfill\eject
\end{document}